\documentstyle[12pt,aps,floats,bezier,epsf,epsfig,psfig,amssymb]{revtex}

\newcommand{\R}{{\Bbb R}}
\newcommand{\C}{{\Bbb C}}
\def\emline#1#2#3#4#5#6{%
       \put(#1,#2){\special{em:moveto}}
       \put(#4,#5){\special{em:lineto}}}
\def\newpic#1{}
\newcommand{\bB}{{\bf B}}
\newcommand{\bM}{{\bf M}}
\newcommand{\bQ}{{\bf Q}}
\newcommand{\bv}{{\bf v}}

\newcommand{\hI}{\hat{I}}
\newcommand{\hcG}{\hat{\cal G}}
\newcommand{\an}{\alpha}
\newcommand{\bn}{\beta}

\newcommand{\ad}{_{\alpha}}
\newcommand{\ab}{_{\alpha\beta}}

\newcommand{\ajd}{_{\alpha,j}}

\newcommand{\abd}{_{\alpha\beta}}

\newcommand{\Y}{\Upsilon}
\newcommand{\anum}{\alpha=1,2,3,\enspace j=1,2,...,n_{\alpha}}

\newcommand{\rJ}{{\rm J}}
\newcommand{\rJt}{{\rm J}^{\dagger}}
\newcommand{\rJo}{{\rm J}_{0}}
\newcommand{\rJot}{{\rm J}_{0}^{\dagger}}
\newcommand{\Om}{{\Omega}}
\newcommand{\Omt}{\Omega^\dagger}

\newcommand{\Psis}{\Psi^{*}}
\newcommand{\Rt}{{\R}^{3}}

\newcommand{\Sum}{\sum\limits}
\newcommand{\opla}{\mathop{\oplus}\limits}
\newcommand{\reduction}[2]{\left. #1 \right|_{#2}}
\newcommand{\diag}{{\rm diag}}

\newcommand{\Real}{\mathop{\rm Re}}
\newcommand{\be}{\begin{equation}}
\newcommand{\ee}{\end{equation}}
\newcommand{\Frac}[2]{\frac{\textstyle #1}{\textstyle #2}}
\newcommand{\D}{\displaystyle}
\begin{document}
\title{Using Faddeev Differential Equations to
Calculate Three-Body Resonances} 
\author{E. A. Kolganova and  A. K. Motovilov} 
\address{Joint Institute  for Nuclear Research, Dubna, 141980, Russia} 
\maketitle 

\bigskip
\small

\noindent Algorithm, based on explicit representations for analytic
continuation of the T-matrix Faddeev components on unphysical
sheets, is worked out for calculations of resonances in the
three-body quantum problem. According to the representations,
poles of T--matrix, scattering matrix and Green function on
unphysical sheets, interpreted as resonances, coincide with those
complex energy values where appropriate truncations of the
scattering matrix have zero as eigenvalue.  Scattering
amplitudes on the physical sheet, necessary to construct
scattering matrix, are calculated on the basis of the Faddeev
differential equations.  The algorithm developed is applied to
search for the resonances in the $nnp$ system and in a model
three-boson system.
\normalsize 

\medskip
 
    LANL E-print {\tt nucl-th/9602001}

    Published in Phys. Atom. Nucl. {\bf 60} No.~2 (1997), 177--185 

\bigskip 

\bigskip

\section{Introduction}
\label{SIntro}
Resonances are one of most interesting phenomena in scattering
of quantum particles. The problem of definition and studying
resonances is payed a lot of attention both in physical and
mathematical literature (see,~e.~g., the books~\cite{Baz} --
\cite{AlbeverioBook}). Main difficulties connected with a
rigorous definition of resonance are explicitly emphasized by
B.~Simon in his survey~\cite{SimonChem}.  A presence of such
difficulties is obliged, first of all, to the fact that in a
contrast to the usual spectrum, resonances are not an unitary
invariant of an operator (Hamiltonian of a quantum system).
The generally accepted interpretation of resonance as a
complex pole of the scattering matrix continued analytically
on unphysical sheet(s) of the energy plane, goes back to the
known paper by G.~Gamow~\cite{GGamow}. For radially symmetric
potentials, such an interpretation of the two-body resonances
has been rigorously approved by R.~Jost~\cite{Jost}.
Beginning from E.C.~Titchmarsh~\cite{Titchmarsh}
resonances are considered as well as poles of analytic continuation
of the Green function (or its matrix elements between suitable
states~\cite{ReedSimonIII}, \cite{ReedSimonIV}).  A survey
of different physical approaches to studying three-body resonances
may be found e.~g., in~\cite{Kukulin}
and~\cite{Orlov}.

At the moment, one of the most effective approaches to practical
calculation of resonances is the complex scaling
method~\cite{BalslevCombes} (see also~\cite{ReedSimonIV},
\cite{SimonChem}).  This method is applicable to the few-body
problem in the case where interaction potentials between
particles are analytic functions of coordinates.  The complex
scaling gives a possibility to rotate the continuous spectrum of
the $N$-body Hamiltonian in such a way that certain sectors
become accessible for observation on unphysical sheets
neighboring with the physical one.  At the same time, the real
discrete spectrum of the Hamiltonian stays fixed during all the
scaling transformation.  Resonances in the sectors above turn
out to be extra discrete eigenvalues of the scaled
Hamiltonian~\cite{ReedSimonIV}.
Thereby, when searching for resonances one may use standard
methods to find discrete spectrum.  Practical applications of
the complex scaling method to concrete problems may be found, in
particular, in the recent papers~\cite{Hu} -- \cite{Szoto}.
Alongside with the complex scaling, another methods are
used for calculations of three-body resonances which are
based in particular on solving the momentum space Faddeev integral
equations~\cite{Faddeev63}, \cite{MF} continued through the cut
(see the survey~\cite{Orlov} by K.~M\"oller and Yu.V.~Orlov and
the literature cited therein). In this approach, resonances are
searched for as poles of the T-matrix.

The present paper is devoted to developing a method to calculate
three-body resonances using the recently found explicit
representations disclosing a structure of the T-matrix on
unphysical sheets as well as analogous representations for the
scattering matrix and resolvent~\cite{MotFewBodyCEBAF},
\cite{MotRemJINR}. These representations were obtained in
supposition that the interaction potentials were pairwise
and falling--off in the coordinate representation not slower than
exponentially.  According to the
representations~\cite{MotFewBodyCEBAF}, \cite{MotRemJINR}, the
matrix $\bM(z)=\{ M\abd(z)\}$, $\alpha,\beta=1,2,3,$ constructed
of the operator $T(z)$ Faddeev components~\cite{Faddeev63},
\cite{MF}, is explicitly expressed on unphysical sheet $\Pi_l$ of
the energy $z$ plane in terms of this matrix itself taken on the
physical sheet and a certain truncation $S_l(z)$  of the total
three-body scattering matrix $S(z)$. Character of the truncation
is determined by the index (number) $l$ of the unphysical sheet
concerned.  Respective representations for analytic continuation
of the matrix $S(z)$ and resolvent $R(z)$ follow immediately
from the representations for $\reduction{\bM(z)}{\Pi_l}$.  A main
consequence of the representations admitting direct practical
applications, is the fact that the T-matrix and resolvent as well as
the scattering matrix have nontrivial singularities on
unphysical sheet $\Pi_l$ exactly at those values of the energy $z$
where the corresponding matrix $S_l(z)$ has zero as eigenvalue.
It is important that $S_l(z)$ is considered on the physical
sheet only. Therefore, one can provide a search for resonances
(poles of $\bM(z)$, $S(z)$ and $R(z)$~) on a certain unphysical
sheet $\Pi_l$ keeping $z$ always on the physical one and calculating
only a position of zeros of the operator-valued function
$S_l(z)$.  For all this, one can use any method allowing
to calculate (on the physical sheet) amplitudes of the processes
necessary to construct the truncation $S_l(z)$.

In the present paper, the matrices $S_l(z)$ are
computed on the base of the numerical algorithm~\cite{MGL} elaborated
to solve the Faddeev differential equations in
configuration space (see the book~\cite{MF},
survey~\cite{EChAYa} and references therein).  Certainly, when
computing the amplitudes on the physical sheet one has to extend
the Faddeev differential formulation of the scattering problem
as well on the complex values of $z$. It should be noted that, in
the holomorphy domain (see~\cite{MotRemJINR}) of the amplitudes,
the differential formulation stays to be correct.

Unfortunately, the algorithm~\cite{MF}, \cite{MGL},
\cite{EChAYa} (see also~\cite{BGC} --\cite{YakovlevFilikhin}) has been
worked out in details only for the processes $(2\longrightarrow
2,3)$. Thus, there may be computed in practice only the
amplitudes of elastic scattering and rearrangement for the
processes $(2\longrightarrow 2)$ and the breakup amplitude
into three particles.  A knowledge of these amplitudes is
sufficient to compute those truncations $S_l(z)$ of the
three-body scattering matrix  $S(z)$, zeros of which are
``responsible'' for resonances situated on the so-called
two-body unphysical sheets, i.~e. those sheets of the energy Riemann
surface where the parameter $z$ may be guided going around the
pair thresholds only.
As a concrete application of the method concerned we make a
search for resonances in the $nnp$ system and in a model system
of three bosons with the nucleon masses.

Let us describe shortly structure of the paper.

In Sec.~\ref{repres}, we introduce main notations and formulate
the explicit representations~\cite{MotRemJINR} for the
unphysical-sheet three-body  T-matrix, scattering matrix and
resolvent which are used then to approve the numerical
method of the work.

In Sec.~\ref{results} the system $nnp$ and a three-boson system
are considered. Formulations of the boundary-value
problems~\cite{MF}, \cite{MGL}, \cite{EChAYa}  are given for the
Faddeev partial differential equations, corresponding to the
processes $(2\longrightarrow 2,3)$ in these systems and going
out to a domain of complex energy values in the physical sheet.
Numerical method to solve these problems is described.
Truncated (partial) scattering matrices are constructed in terms
of the amplitudes for elastic scattering $(2\longrightarrow2)$.
Zeros of these matrices represent resonances (including virtual poles)
on the unphysical sheet connected with the physical one by
crossing the continuous spectrum interval between the deuteron
energy and three-body threshold.  Results of numerical
computations are exposed.

We conclude this introduction with some notation. Throughout the paper,
we understand by $\sqrt{z-\lambda}$,
$z\in{\C}$, $\lambda\in{\R}$, the main branch of the function
$(z-\lambda)^{1/2}$.  By $\hat{p}$ we denote the unit vector
in the direction
$p\in{\R}^n$, $\hat{p}=\Frac{p}{|p|}$, and by $S^{n-1}$,
the unit sphere in ${\R}^n$, $\hat{p}\in S^{n-1}$.

\section{ Explicit representations for '-matrix, scattering matrix
and resolvent on unphysical sheets}
\label{repres}
The scattering matrix,  '-matrix and Hamiltonian resolvent (Green
function) for a quantum-mechanical system are stringently
connected with each other.  Therefore, all these three objects,
considered as functions of energy, have usually the same Riemann
surface.  Such a fact takes place at least in the multichannel
scattering problem with binary channels and in the three-body problem with
quickly decreasing interactions~\cite{TMF93},
\cite{MotRemJINR}.  In a ``first approximation'', the structure of
the Riemann surfaces in these problems coincides.
The thing is that the branching points
(in real axis) are stipulated to these surfaces in the both problems
by the same reason, namely by a presence of the Cauchy-type
integrals in the Lippmann--Schwinger or Faddeev equations.
In the equations considered in the momentum representation,
the Cauchy-type integrals are engendered by the kernels
$
\Frac{\delta(p-p')}{\lambda+p^{2}-z}
$
with $\lambda$, the channel thresholds and $p$, $p\in
{\R}^{n}$, the respective channel
momentum variables. In the case of the
channels $(2\longrightarrow 2,3)$ in three-body problem and the
odd-dimensional channels (i.~e., with $n$ odd) in the matrix
multichannel problem, the thresholds $\lambda$ turn out to be
the branching points of the second order.  Even-dimensional
channels in the multichannel problem as well as the channel
$(3\rightarrow 2,3)$ in the three-body problem give logarithmic
branching points (see~\cite{MotRemJINR},
\cite{TMF93}, \cite{Orlov}).

The method used for calculation of resonances in the present work,
is based on the explicit representations~\cite{MotRemJINR} for
analytic continuation of the T-matrix, scattering matrices and
Green function on unphysical sheets keeping true at least for a
part of the three-body Riemann surface.
To describe this part we introduce the auxiliary vector-function
$f(z)=(f_{0}(z)),f_{1,1}(z),..., f_{1,n_1}(z),$
$f_{2,1}(z),...,f_{2,n_2}(z),$ $f_{3,1}(z),...,f_{3,n_3}(z))$
$\enspace$
with $\enspace  f_{0}(z)$ $ =$ $\ln{z}$ and $f\ajd(z)$ $=$
$(z-\lambda\ajd)^{1/2}.$ Here, by $\lambda\ajd$ we understand
respective bound-state energies of pair subsystems $\an,$
$\an=1,2,3,$ $j=1,2,...,n\ad,$ numerated taking into account
their multiplicities.  It is supposed that $n\ad<\infty$.  The
sheets $\Pi_{l}$ of the vector-function $f(z)$ Riemann surface
$\Re$ are numerated via multi-index
$
l=( l_{0},l_{1,1},..., l_{1,n_1},
l_{2,1},...,l_{2,n_2},
l_{3,1},...,l_{3,n_3}),
$
where $l\ajd=0$ if the sheet $\Pi_{l}$ corresponds to the main
(arithmetic) branch of the square root $(z-\lambda\ad)^{1/2}.$
Otherwise, $l\ajd=1$ is assumed.  Value of $l_0$ coincides with
the number of branch of the function $\ln{z}$, $\ln z=\ln|z|+i\,2\pi
l_0+i\phi$ where $\phi=\mathop{\rm arg}z$. For the physical
sheet identified by $l_0=l\ajd=0$, $\anum$, we use the notation
$\Pi_0$.  A ``sticking'' of the sheets $\Pi_l$ of the surface
$\Re$ is realized in intervals between neighboring thresholds
along rims of the cut along the continuous spectrum.  A detailed
description of the surface $\Re$ may be found in~\cite{TMF93}.

The type $\Re$ surfaces without extra branching points arise in
the multichannel problems with binary channels only.  Structure
of the total three-body Riemann surface is essentially more
complicated. For instance, the sheets $\Pi_{l}$ with $l_0=\pm 1$
have additional branching points corresponding to resonances of
the two-body subsystems. In the sheets $\Pi_l$ with $l_0=0$, one can
discover (in the left half-plane) logarithmic branching points
of a kinematical origin.  The part $\Re^{(3)}$ of the total
three-body Riemann surface where the
representations~\cite{MotRemJINR} are valid, consists of the
surface $\Re$ sheets $\Pi_l$ identified by $l_0=0$ (such unphysical
sheets are called {\it two-body} sheets) and two {\it
three-body} sheets identified by $l_0=\pm 1$ and $l\ajd=1,$
$\an=1,2,3,$ $j=1,2,...,n\ad.$  Note that $\Re^{(3)}$ includes
all the unphysical sheets neighboring with the physical sheet
 $\Pi_0$.

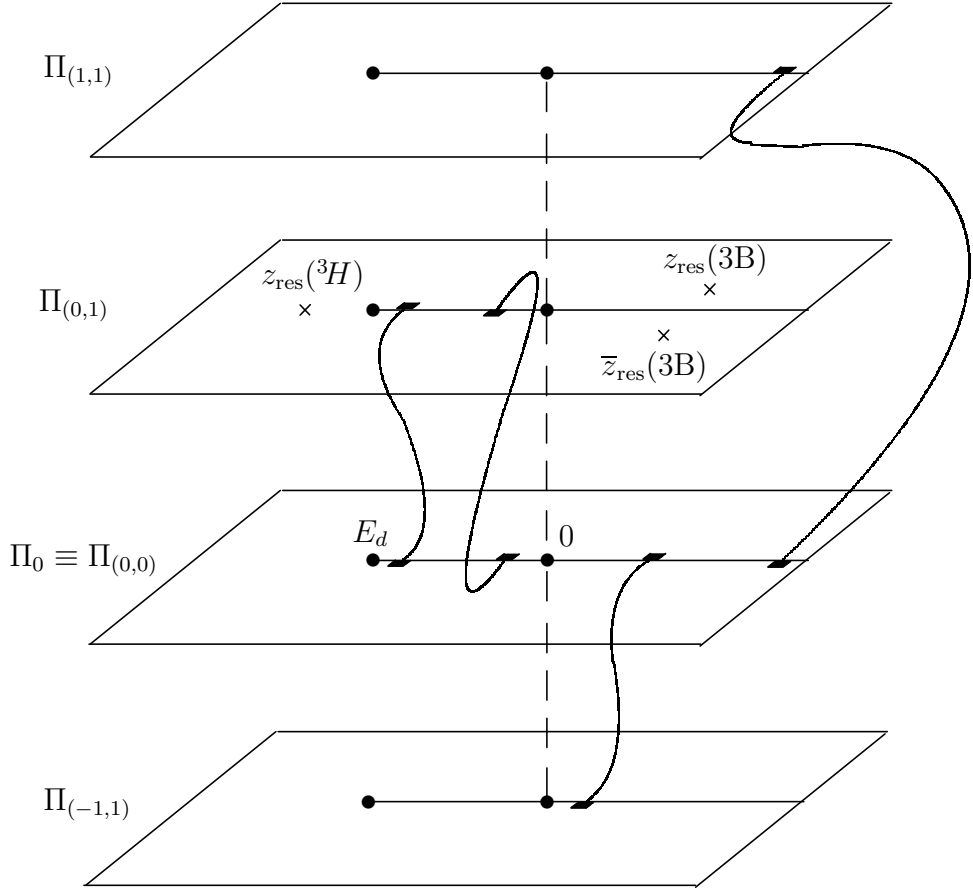
\begin{figure}
\centering
\unitlength=0.9mm
\special{em:linewidth .6pt}
\linethickness{.6pt}
\begin{picture}(161.48,162.67)
\emline{19.67}{32.33}{1}{47.67}{55.00}{2}
\emline{109.67}{32.00}{3}{137.67}{54.67}{4}
\emline{19.33}{32.29}{5}{109.67}{32.29}{6}
\emline{48.00}{55.00}{7}{138.00}{55.00}{8}
\emline{61.33}{44.67}{9}{125.67}{44.67}{10}
\put(61.36,44.68){\circle*{1.92}}
\put(87.76,44.68){\circle*{1.92}}
\emline{20.33}{68.00}{11}{48.33}{90.67}{12}
\emline{110.33}{67.67}{13}{138.33}{90.33}{14}
\emline{20.09}{67.95}{15}{110.42}{67.95}{16}
\emline{48.67}{90.67}{17}{138.67}{90.67}{18}
\emline{62.00}{80.33}{19}{126.33}{80.33}{20}
\put(62.02,80.35){\circle*{1.92}}
\put(87.82,80.48){\circle*{1.92}}
\emline{20.33}{105.00}{21}{48.33}{127.67}{22}
\emline{110.33}{104.67}{23}{138.33}{127.33}{24}
\emline{20.20}{104.94}{25}{110.53}{104.94}{26}
\emline{48.67}{127.67}{27}{138.67}{127.67}{28}
\emline{62.00}{117.33}{29}{126.33}{117.33}{30}
\put(62.02,117.35){\circle*{1.92}}
\put(87.77,117.35){\circle*{1.92}}
\emline{20.33}{140.00}{31}{48.33}{162.67}{32}
\emline{110.33}{139.67}{33}{138.33}{162.33}{34}
\emline{20.30}{139.96}{35}{110.63}{139.96}{36}
\emline{48.67}{162.67}{37}{138.67}{162.67}{38}
\emline{62.00}{152.33}{39}{126.33}{152.33}{40}
\put(62.02,152.35){\circle*{1.92}}
\put(87.77,152.35){\circle*{1.92}}
\emline{87.67}{146.67}{41}{87.67}{151.00}{42}
\emline{87.67}{139.00}{43}{87.67}{143.33}{44}
\emline{87.67}{132.00}{45}{87.67}{136.33}{46}
\emline{87.67}{125.00}{47}{87.67}{129.33}{48}
\emline{87.67}{117.67}{49}{87.67}{122.00}{50}
\emline{87.67}{112.33}{51}{87.67}{116.67}{52}
\emline{87.67}{104.67}{53}{87.67}{109.00}{54}
\emline{87.67}{97.67}{55}{87.67}{102.00}{56}
\emline{87.67}{90.67}{57}{87.67}{95.00}{58}
\emline{87.67}{83.33}{59}{87.67}{87.67}{60}
\emline{87.73}{74.33}{61}{87.73}{78.67}{62}
\emline{87.73}{66.67}{63}{87.73}{71.00}{64}
\emline{87.73}{59.67}{65}{87.73}{64.00}{66}
\emline{87.73}{52.67}{67}{87.73}{57.00}{68}
\emline{87.73}{45.33}{69}{87.73}{49.67}{70}
\emline{123.45}{152.42}{71}{124.52}{153.24}{72}
\emline{121.23}{152.45}{73}{122.30}{153.26}{74}
\emline{121.21}{152.40}{75}{123.42}{152.40}{76}
\emline{122.30}{153.18}{77}{124.49}{153.18}{78}
\emline{122.10}{153.06}{79}{124.26}{153.06}{80}
\emline{121.94}{152.93}{81}{124.11}{152.93}{82}
\emline{121.79}{152.80}{83}{123.96}{152.80}{84}
\emline{121.64}{152.70}{85}{123.83}{152.70}{86}
\emline{121.46}{152.60}{87}{123.65}{152.60}{88}
\emline{121.33}{152.50}{89}{123.52}{152.50}{90}
\emline{93.52}{43.74}{91}{94.58}{44.56}{92}
\emline{91.30}{43.77}{93}{92.37}{44.58}{94}
\emline{91.28}{43.72}{95}{93.49}{43.72}{96}
\emline{92.37}{44.51}{97}{94.56}{44.51}{98}
\emline{92.17}{44.38}{99}{94.33}{44.38}{100}
\emline{92.01}{44.25}{101}{94.18}{44.25}{102}
\emline{91.86}{44.12}{103}{94.02}{44.12}{104}
\emline{91.71}{44.02}{105}{93.90}{44.02}{106}
\emline{91.53}{43.92}{107}{93.72}{43.92}{108}
\emline{91.40}{43.82}{109}{93.59}{43.82}{110}
\emline{104.17}{80.52}{111}{105.24}{81.34}{112}
\emline{101.95}{80.55}{113}{103.02}{81.36}{114}
\emline{101.93}{80.50}{115}{104.14}{80.50}{116}
\emline{103.02}{81.28}{117}{105.21}{81.28}{118}
\emline{102.82}{81.16}{119}{104.98}{81.16}{120}
\emline{102.67}{81.03}{121}{104.83}{81.03}{122}
\emline{102.51}{80.90}{123}{104.68}{80.90}{124}
\emline{102.36}{80.80}{125}{104.55}{80.80}{126}
\emline{102.18}{80.70}{127}{104.37}{80.70}{128}
\emline{102.06}{80.60}{129}{104.24}{80.60}{130}
\emline{122.61}{79.36}{131}{123.68}{80.17}{132}
\emline{120.40}{79.38}{133}{121.47}{80.20}{134}
\emline{120.37}{79.33}{135}{122.59}{79.33}{136}
\emline{121.47}{80.12}{137}{123.66}{80.12}{138}
\emline{121.26}{79.99}{139}{123.43}{79.99}{140}
\emline{121.11}{79.87}{141}{123.27}{79.87}{142}
\emline{120.96}{79.74}{143}{123.12}{79.74}{144}
\emline{120.80}{79.64}{145}{122.99}{79.64}{146}
\emline{120.63}{79.54}{147}{122.82}{79.54}{148}
\emline{120.50}{79.43}{149}{122.69}{79.43}{150}
\emline{82.54}{80.47}{151}{83.61}{81.28}{152}
\emline{80.33}{80.50}{153}{81.40}{81.31}{154}
\emline{80.30}{80.44}{155}{82.52}{80.44}{156}
\emline{81.40}{81.23}{157}{83.59}{81.23}{158}
\emline{81.20}{81.11}{159}{83.36}{81.11}{160}
\emline{81.04}{80.98}{161}{83.21}{80.98}{162}
\emline{80.89}{80.85}{163}{83.05}{80.85}{164}
\emline{80.74}{80.75}{165}{82.93}{80.75}{166}
\emline{80.56}{80.65}{167}{82.75}{80.65}{168}
\emline{80.43}{80.55}{169}{82.62}{80.55}{170}
\emline{66.49}{79.52}{171}{67.56}{80.33}{172}
\emline{64.27}{79.54}{173}{65.34}{80.36}{174}
\emline{64.25}{79.49}{175}{66.46}{79.49}{176}
\emline{65.34}{80.28}{177}{67.53}{80.28}{178}
\emline{65.14}{80.15}{179}{67.30}{80.15}{180}
\emline{64.98}{80.03}{181}{67.15}{80.03}{182}
\emline{64.83}{79.90}{183}{67.00}{79.90}{184}
\emline{64.68}{79.80}{185}{66.87}{79.80}{186}
\emline{64.50}{79.69}{187}{66.69}{79.69}{188}
\emline{64.37}{79.59}{189}{66.56}{79.59}{190}
\emline{67.82}{117.58}{191}{68.89}{118.39}{192}
\emline{65.61}{117.60}{193}{66.68}{118.42}{194}
\emline{65.58}{117.55}{195}{67.80}{117.55}{196}
\emline{66.68}{118.34}{197}{68.87}{118.34}{198}
\emline{66.47}{118.21}{199}{68.64}{118.21}{200}
\emline{66.32}{118.09}{201}{68.48}{118.09}{202}
\emline{66.17}{117.96}{203}{68.33}{117.96}{204}
\emline{66.01}{117.86}{205}{68.20}{117.86}{206}
\emline{65.84}{117.76}{207}{68.03}{117.76}{208}
\emline{65.71}{117.65}{209}{67.90}{117.65}{210}
\emline{80.63}{116.53}{211}{81.70}{117.34}{212}
\emline{78.41}{116.55}{213}{79.48}{117.37}{214}
\emline{78.39}{116.50}{215}{80.60}{116.50}{216}
\emline{79.48}{117.29}{217}{81.67}{117.29}{218}
\emline{79.28}{117.16}{219}{81.44}{117.16}{220}
\emline{79.12}{117.04}{221}{81.29}{117.04}{222}
\emline{78.97}{116.91}{223}{81.14}{116.91}{224}
\emline{78.82}{116.81}{225}{81.01}{116.81}{226}
\emline{78.64}{116.71}{227}{80.83}{116.71}{228}
\emline{78.51}{116.60}{229}{80.70}{116.60}{230}
\bezier{100}(66.40,80.44)(72.77,84.86)(66.40,101.09)
\bezier{88}(66.51,117.43)(59.21,111.98)(66.40,101.09)
\bezier{224}(81.22,80.16)(70.28,65.21)(80.69,100.71)
\bezier{224}(80.18,117.30)(92.06,134.02)(80.64,100.70)
\bezier{96}(93.39,44.62)(100.17,49.48)(97.48,65.21)
\bezier{76}(97.36,65.21)(96.08,76.34)(102.60,80.30)
\bezier{316}(122.51,80.46)(161.48,117.64)(145.29,136.62)
\bezier{92}(125.51,141.62)(139.50,143.22)(145.29,136.62)
\bezier{92}(122.42,152.15)(110.08,142.24)(117.71,142.08)
\bezier{28}(117.88,141.75)(122.26,141.59)(125.35,141.43)
\put(8.33,80.33){\makebox(0,0)[lc]{$\Pi_0\equiv\Pi_{(0,0)}$}}
\put(12.66,118.00){\makebox(0,0)[lc]{$\Pi_{(0,1)}$}}
\put(13.33,44.33){\makebox(0,0)[lc]{$\Pi_{(-1,1)}$}}
\put(13.33,152.66){\makebox(0,0)[lc]{$\Pi_{(1,1)}$}}
\emline{51.28}{116.56}{231}{52.75}{118.14}{232}
\emline{51.25}{118.08}{233}{52.68}{116.65}{234}
\emline{111.06}{119.60}{235}{112.54}{121.18}{236}
\emline{111.04}{121.12}{237}{112.47}{119.69}{238}
\emline{104.25}{112.81}{239}{105.72}{114.39}{240}
\emline{104.22}{114.32}{241}{105.65}{112.89}{242}
\put(53.12,122.47){\makebox(0,0)[cc]{$z_{\rm res}({}^3\!{H})$}}
\put(61.67,84.33){\makebox(0,0)[cc]{$E_d$}}
\put(90.67,84.00){\makebox(0,0)[cc]{0}}
\put(112.59,124.66){\makebox(0,0)[cc]{$z_{\rm res}(\mbox{3B})$}}
\put(103.59,108.93){\makebox(0,0)[cc]{$\overline{z}_{\rm res}(\mbox{3B})$}}
\end{picture}
\caption{ Physical, $\Pi_0$, and neighboring unphysical,
$\Pi_{(0,1)}$, $\Pi_{(1,1)}$ and $\Pi_{(-1,1)}$,
sheets of the Riemann surface in the three-nucleon
$(nnp)$ and model three-boson problems.
The near (in figure) rim of the cut in the sheet $\Pi_0$
is identified in the interval between the deuteron energy
$E_d$ and three-body threshold
$z=0$ with the remote rim of the cut in the sheet
$\Pi_{(0,1)}$. On the contrary, the remote rim in the sheet
$\Pi_0$ is identified in this interval with the near
rim in the sheet
$\Pi_{(0,1)}$. In the interval $(0,+\infty)$,
one identifies the near rim of the cut in the sheet $\Pi_0$
with the remote rim in the sheet $\Pi_{(1,1)}$. Respectively,
the remote rim in the sheet $\Pi_0$ is identified in this interval
with the near rim in the sheet $\Pi_{(-1,1)}$. On the sheet $\Pi_{(0,1)}$,
the virtual level $z_{\rm res}({}^3\!{H})$ in the s-state of the
$nnp$-system is marked, and the resonance $z_{\rm res}(\mbox{3B})$
in the model three-boson system. Also, the complex conjugate
pole  $\overline{z}_{\rm res}(\mbox{3B})$ is marked.
}
\label{RimSurf-nnp}
\end{figure}

In the case of the $nnp$-system and a model three-boson system
considered below, the surface $\Re^{(3)}$ is shown schematically
in Fig.~\ref{RimSurf-nnp}. In this case a single pair threshold,
the deuteron energy $\lambda=E_d$ is present only. Therefore, the
index $l$ of the sheets  $\Pi_l$ consists of two components
only:  $l=(l_0,l_1)$. In the terminology accepted, the sheet
$\Pi_{(0,1)}$ (see Fig.~\ref{RimSurf-nnp}) is a two-body sheet, but the
sheets $\Pi_{(-1,1)}$ and $\Pi_{(1,1)}$ are three-body ones.

Construction of the representations for the T-matrix consists of the
following stages.  At the first step, one carries out analytic
continuation on unphysical sheets, of the absolute terms and
kernels of the Faddeev integral equations for the components
$M\abd(z)$ (the continuation is understood in the sense of
distributions).  As the absolute terms as the kernels after
continuation are expressed in terms of the pair T-matrices and
scattering matrices taken on the physical sheet.  Transforming
the Faddeev equations continued, one finds the kernels
$
\reduction{M\ab (P,P',z)}{z\in\Pi_l}
$
can be explicitly expressed in terms of these kernels themselves
taken on the physical sheet $\Pi_0$ in their off-shell and/or
half-on-shell variants.  It is supposed in the last case that
the first argument $P$ of the kernels
$
\reduction{M\ab (P,P',z)}{z\in\Pi_l}
$
is taken on the energy (``mass'') shells
$|P|^{2}=z$ or
$|p\ad|^{2}=z-\lambda\ajd,$ $j=1,2,...,n\ad.$ Here,
we use the notations
$P=\{k\ad, p\ad\}$ with $k\ad, p\ad,$
$\an=1,2,3,$ standing for the standard relative momenta~\cite{MF}.
Transferring in the expressions obtained all the off-shell terms
to the l.h.~part  and inverting an operator arising there,
one comes to a closed system of equations for the half-on-shell
components $\reduction{M\ab (P,P',z)}{z\in\Pi_l}$.
This system admits an explicit solution using the terms
of the physical sheet only.
As a result one gets the following representations%
\footnote{For the sake of simplicity, we write here these
representations as well as representations for the scattering
matrix and resolvent (see below Eqs.~(\ref{SS}) and~(\ref{RR}),
respectively) for the case of spinless particles only.  A direct
generalization of the representations on the case of spin
particles causes no difficulties.}
for the matrix
$
\bM(z)=\left\{ M\ab(z) \right\}, \an,\bn=1,2,3,
$
continued on the sheet $\Pi_l$:
\be
\label{Mab}
\reduction{\bM(z)}{\Pi_l}=
\bM(z) - \bB^{\dagger}(z) A(z)L S_{l}^{-1}(z)\tilde{L} \bB(z).
\ee
Here, the factor $A(z)$ is the diagonal matrix,
$$
A(z)=\diag\{ A_{0}(z), A_{1,1}(z),...,A_{1,n_3}(z) \},
$$
combined of the functions
$
A_{0}(z)=-\pi i z^2
$
and
$
A\ajd=-\pi i \sqrt{z-\lambda\ajd}.
$
Notations  $L$ and $\tilde{L}$ are used for diagonal
number matrices whose nontrivial elements are the sheet
$\Pi_l$ indices:
$$
L=\diag\{ l_0 , l_{1,1},...,l_{3,n_3}\} \,\, \mbox{\rm and} \,\,\,
\tilde{L}=\diag\{ |l_0| , l_{1,1},...,l_{3,n_3}\}.
$$
By $S_l(z)$ we understand a truncation
of the three-body scattering matrix:
$S(z)$,\,\,
$S(z):\, \hcG\rightarrow \hcG,$ \,\,
$\hcG=L_2(S^5)\opla_{\an=1}^{3} \opla_{j=1}^{n\ad} L_2(S^2),$
defined by the equation

$$
S_l (z)= \hat{I} + \tilde{L}\bigl[  S(z)-\hat{I}  \bigr] L
$$
where $\hI$ is the identity operator in  $\hcG$.
Also, we use the notations

$$
\bB(z)=\left( \begin{array}{l}
     \rJo\Om \bM    \\
     \rJ_1\Psis [\Y \bM+\bv]
    \end{array}
\right)
\mbox{\,\, and \,\,}
\bB^{\dagger}(z)=\left( \bM(z)\Omt\rJot,
[\bv +\bM\Y ]\Psi \rJt_1\right).
$$
Here,
$
    \bv=\diag\{ v_1 , v_2 , v _3 \}
$
with $v_\alpha$, the pair potentials, $\alpha=1,2,3$. At the same time,
$$
 \Om=(1,\,\, 1,\,\, 1), \quad
\Y= \left(\begin{array}{lcr}
0     &    1   &     1\\
1     &     0  &     1 \\
1     &     1  &     0
\end{array}\right)
\mbox{\,\, and\,\, }
\Psi=\diag\{ \Psi_1 , \Psi_2 , \Psi_3 \}
$$
where $\Psi\ad,$ $\an=1,2,3,$ are operators acting on
$
f=(f_1,f_2,...,f_{n\ad})\in \opla_{j=1}^{n\ad} L_2 (\Rt)
$
as

$$
(\Psi\ad f)(P)=\Sum_{j=1}^{n\ad}\psi\ajd(k\ad)f_{j}(p\ad),
$$
where, in turn,  $\psi\ajd$ is the bound
state  wave function of the pair subsystem $\an$
corresponding to the level
$\lambda_{\alpha,j}$.  By $\Psis$ we denote operator adjoint to
$\Psi$.  Notation $\rJo(z)$ is used for operator restricting a
function on the energy-shell $|P|^2=z$.  The diagonal
matrix-valued function

$$
   \rJ_1(z)=\diag\{ \rJ_{1,1}(z),...,\rJ_{3,n_3} (z)\},
$$
consists of the operators ${\rm J}\ajd(z)$ of restriction on the
energy surfaces
$
|p\ad|^{2}=z-\lambda\ajd.
$
The operators $\Omt$,  $\rJot(z)$ and $\rJt_1(z)$ represent the
``transposed'' matrices $\Om$, $\rJo(z)$ and $\rJ_1(z)$,
respectively.  Operators  $\rJot(z)$ and $\rJt_1(z)$  act in the
expression for $\bB^{\dagger}$ (as if) to the left.

Representations for the scattering matrix and resolvent on
unphysical sheets are an immediate consequence of the
representations~(\ref{Mab}) for the matrix
$\reduction{\bM(z)}{\Pi_l}$.

With details omitted (see~\cite{MotRemJINR}) the
representations for the scattering matrix $S(z)$  read 
\be
\label{SS}
\reduction{S(z)}{\Pi_l}={\cal E}(l)\left\{ \hat{I} +
S_{l}^{-1}(z) [S(z)-\hat{I}]e(l)\right\} {\cal E}(l).
\ee
Here, $ {\cal E}=\diag\{{\cal E}_0 ,{\cal E}_{1,1},..., {\cal
E}_{3,n_3} \} $ where $ {\cal E}_0 $ is the identity operator in
$L_2 (S^5 )$ if  $l_0 =0$ and ${\cal E}_0$, the inversion, $({\cal
E}_0 f)(\hat{P}) = f(-\hat{P}),$ if $l_0 =\pm 1$.  Analogously,
$ {\cal E}\ajd $ is the identity operator in $L_2 (S^2 )$ for $
l\ajd =0 $ and inversion for $ l\ajd =1.$
Notation $e(l)$ is used for diagonal number matrix
$e(l)=\diag\{e_0 ,e_{1,1},...,e_{3,n_3} \}$ with nontrivial
elements $e\ajd=1$ if $l\ajd=0$ and $e\ajd=-1$ if $l\ajd=1;$ for
all the cases $e_0=1$.

Analytic continuation $\reduction{R(z)}{\Pi_l}$
of the resolvent $R(z)=(H-z)^{-1}$ of the three-body Hamiltonian
$H$ admits the representation
\begin{equation}
\label{RR}
 \reduction{R(z)}{\Pi_l}=R +\bQ^\dagger AL S_l^{-1}\tilde{L}\bQ.
\end{equation}
Here,
$$
\bQ= \left(
\begin{array}{c}  \rJo[I-VR]   \\     \rJ_1 \Psis[I-\Y\bM R_0]\Omt
\end{array} \right), \quad
\bQ^\dagger=\bigl( [I-RV]\rJot\, ,\,\,
\Om[I-R_0\bM\Y]\Psi\rJt_1 \bigr)
$$
with $V=v_1+v_2+v_3$ and $R_0(z)=(H_0-z)^{-1}$, the resolvent of
the kinetic energy operator $H_0$ for the system under
consideration.

There were holomorphy domains $\Pi_l^{({\rm hol})}$ found
in~\cite{MotRemJINR} for the truncated scattering matrices
$S_l(z)$ in the physical sheet.  Representations~(\ref{Mab})
and~(\ref{RR}) are valid in the same domains.

It follows from the representations~(\ref{Mab})---({\ref{RR})
that the resonances (the nontrivial poles of
$\reduction{\bM(z)}{\Pi_l},$ $\reduction{S(z)}{\Pi_l}$ and
$\reduction{R(z)}{\Pi_l}$) situated on the unphysical sheet
$\Pi_l$ are in fact those points $z=z_{\rm res}$ on the physical
sheet where the matrix $S_l(z)$ has zero as eigenvalue.
Therefore, {\em a calculation of resonances on the unphysical
sheet $\Pi_l$ is reduced to a search for zeros of the truncation
$S_l(z)$ of the total three-body scattering matrix $S(z)$ in the
physical sheet}.  At the numerical search for the resonances, one
can use any method allowing to find analytic continuation on the
physical sheet, of the elastic scattering, rearrangement or
breakup amplitudes necessary to construct the respective
truncation ${S}_l(z)$.

\section{Results of search for resonances in three-nucleon
(\lowercase{\bf nnp}) and three-boson systems}
\label{results}
One of the most effective methods for a study of concrete
three-particle systems is the numerical algorithm ~\cite{MGL}
(see also ~\cite{EChAYa},  \cite{MF} and references therein)
based on the Faddeev differential equations for the wave function
components in the configuration space.  This method gives
opportunity of comparatively easy calculations of the scattering
wave functions and respective amplitudes for $(2\rightarrow
2,3)$ processes. An extension of the differential formulation to
a domain of complex energies enables us to calculate analytic
continuation of the $(2\longrightarrow 2,3)$ amplitudes on the
physical sheet.  This is quite enough for us to construct the
truncated s-state scattering matrices $S_{l}(z)$ whose zeros are
resonances on the respective {\it two-body} unphysical sheets
$\Pi_l$ with $l_0=0$.

As to a search for resonances on the {\it three-body} unphysical
sheets $\Pi_l$ with $l_0\neq 0$ (in particular on the sheets
$\Pi_{(-1,1)}$ and $\Pi_{(1,1)}$ in the three-nucleon problem), the
situation is much more complicated. To construct $S_l(z)$ in
this case, one has, alongside with $(2\longrightarrow 2)$, to
calculate amplitudes for the processes $(3\longrightarrow 2,3)$
with three asymptotically free particles in an initial state.
Unfortunately, the reliable practical methods for calculation of
the processes $(3\longrightarrow 2,3)$ are not developed so far
even for the real energies. In particular, when using the
differential formulation of the scattering problem, one has from
the very beginning to separate explicitly contributions to the
Faddeev components not only from single-rescattering but also
from double-rescattering processes~\cite{MF}.  In the last case,
one has to take into account explicitly  (see ~\cite{MF}) a
presence of the ``light'' and ``shadow'' zones for the
correspondent waves, and to use the Fresnel integral for
description of intermediate regimes.  Moreover, computations of
the $(3\longrightarrow 2,3)$ amplitudes have to be carried out
for many different directions of the incident momentum $P$. This
circumstance is itself a factor enlarging numerical complexity
as compared with the case of the processes $(2\longrightarrow
2,3)$.

So, in the present work we restrict ourselves with a search for
the $nnp$ system resonances situated on the unphysical sheet
$\Pi_{(0,1)}$ only, connected with the physical one by crossing
the continuum spectrum interval $(E_d,\, 0)$ between the
deuteron energy $z\!=\! E_d$ and breakup threshold
$z\!=\! 0$  (see~Fig.~\ref{RimSurf-nnp}). To
construct the truncated scattering matrix $S_{(0,1)}(z)$ (in
accordance with~(\ref{Mab})--(\ref{RR}) just its zeros represent
resonances on the sheet $\Pi_{(0,1)}$), it suffices to calculate
the elastic $nd$ scattering amplitudes only.

To find these amplitudes we
use the two--dimensional Faddeev integro--differential
equations ~\cite{MF} being a result of the partial and angular
analysis of the Faddeev differential equations.
In addition we make an assumption, rather usual in nuclear
physics, that the nucleons involved interact in the s-state only.
With this assumption the partial equations become exact.
The Faddeev component $U_L^{q}$ corresponding to the
total spin ${\bf S}=3/2$ and
total orbital momentum $ L $, satisfies the equation

\begin{equation}
\label{Fquartet}
(H_L-z)U_L^q(x,y)=V_t(x)\Psi_L^q(x,y)
\end{equation}
with
$
H_L=-\D\frac{\partial^2}{\partial x^2}
    -\D\frac{\partial^2}{\partial y^2}
    +\D\frac{L(L+1)}{y^2},
$
the partial Laplacian and
$x,y$, the absolute values of the Jacobi vectors~\cite{MF}.
Respective partial component
$\Psi_L^q(x,y)$ of the total wave function for the
$nnp$-system is expressed by the function $U_L^q(x,y)$ as

\begin{equation}
\label{FTqconn}
\Psi_L^q(x,y)=U_L^q(x,y)-\D\frac{1}{2}\int\limits_{-1}^1
du\,h^L(x,y,u)\,U_L^q(x',y')
\end{equation}
where
$x'=\sqrt{\Frac{1}{4}\,x^2+\Frac{3}{4}\,y^2-\Frac{\sqrt{3}}{2}\,xyu }$
and
$y'=\sqrt{\Frac{3}{4}\,x^2+\Frac{1}{4}\,y^2+\Frac{\sqrt{3}}{2}\,xyu }.$
The geometric function $h^L$
is given by
$$
h^L=\D \frac{4}{\sqrt{3}}\, \frac{xy}{x'y'}
\left(\frac{-1}{2\sin\theta'}\right)^L
\,\sum\limits_{k=0}^L
\D\frac{L!}{k!(L-k)!} P_k(u)
\left(\sqrt{3}\cos\theta\right)^k
\left(\sin\theta\right)^{L-k}
$$
with $P_k(u)$, the Legendre polynomial of the order $k$ and
$\theta=\mathop{\rm arctg}\D\frac{y}{x}$, $\theta'=\mathop{\rm
arctg}\D\frac{y'}{x'}$.  The factor $V_t(x)$ represents a
triplet part of the nucleon-nucleon interaction potential.

Function $U_L^q(x,y)$ satisfies the boundary conditions
\begin{equation}
\label{BCQ}
\reduction{U_L^q(x,y)}{x=0}=0     \quad\mbox{\rm and}
\reduction{U_L^q(x,y)}{y=0}=0.
\end{equation}
It satisfies also the asymptotical conditions
\begin{eqnarray}
U_L^q(x,y) &
\mathop{\mbox{\large$\sim$}}\limits_{\rho\longrightarrow\infty} &
\psi_d(x)\left[
j_L(\sqrt{z-E_d}\,y)+a_L^q(z)\exp
\left\{i\sqrt{z-E_d}\,y+i\Frac{\pi L}{2} \right\}
\right] + \nonumber\\
\label{AsQ}
 & &
+ A_L^q(z,\theta)\frac{\exp
\left\{i\sqrt{z}\rho+i\Frac{\pi L}{2}\right\} }{\sqrt{\rho}}
\end{eqnarray}
where $j_L$ is the Bessel spherical function of the order $L$
and $\rho=\sqrt{x^2+y^2}$. By $\psi_d(x)$ we denote the deuteron
wave function and by $a_L^d(z)$, the quartet amplitude for
elastic $nd$-scattering in a state with angular momentum $L$.
The function $A_L^q(z,\theta)$ represents the partial Faddeev
component of the breakup amplitude of this system into three
particles.

The Faddeev integro-differential equations
for the doublet $({\bf
S}=1/2)$ $nd$-scattering read as

\begin{equation}
\label{Fdoublet}
(H_L-z)U_{1(2),L}^d(x,y)=V_{t(s)}(x)\Psi_{1(2),L}^d(x,y)
\end{equation}
where $\Psi_L^d$,
$
\Psi_L^d=
\left(
\begin{array}{c}                  \Psi_{1,L}^d\\
                                  \Psi_{2,L}^d
\end{array}
\right)
$,
is expressed by the vector
$
U_L^d=\left( \begin{array}{c} U_{1,L}^d\\
                              U_{2,L}^d
\end{array}\right)
$
as
\begin{equation}
\label{FTdconn}
\Psi_L^q=U_L^d+\D\frac{1}{2}\int\limits_{-1}^1
du\,h^L(u){\cal B}\, U_L^d, \quad
{\cal B}=
\left(
\begin{array}{rr}
  \Frac{1}{4}  & \phantom{\mbox{$\D\int$}}  -\Frac{3}{4} \\
 -\Frac{3}{4}  & \phantom{\mbox{$\D\int$}}   \Frac{1}{4}
\end{array}
\right).
\end{equation}
By $V_s$ we understand a singlet part of the
nucleon-nucleon interaction potential.

The Faddeev partial components $U_{i,L}^d$ satisfy as
$U_L^q$, the boundary conditions

\begin{equation}
\label{BCD}
\reduction{U_{i,L}^d(x,y)} {x=0} =0,     \quad
\reduction{U_{i,L}^d(x,y)} {y=0} =0,
\end{equation}
and have the asymptotics
\begin{eqnarray}
U_{i,L}^d(x,y) &
\mathop{\mbox{\large$\sim$}}\limits_{\rho\longrightarrow\infty} &
\delta_{i1}\,\psi_d(x)\left[
j_L(\sqrt{z-E_d}\,y)+a_L^d(z)\exp
\left\{i\sqrt{z-E_d}\,y+i\Frac{\pi L}{2} \right\}
\right] + \nonumber\\
\label{AsD}
 & &
+ A_{i,L}(z,\theta)\frac{\exp
\left\{i\sqrt{z}\rho+i\Frac{\pi L}{2}\right\} }{\sqrt{\rho}},
\quad i=1,2.
\end{eqnarray}
Here, $a_L^d(z)$ is the doublet elastic $nd$-scattering
amplitude in the state of the $nnp$-system with angular momentum
$L$. At the same time, the functions $A_{1,L}^d$ and $A_{2,L}^d$
represent the partial Faddeev components of the breakup amplitude
for this state.  Remind that the physical breakup amplitudes in
quartet and doublet states are expressed by the amplitudes
respectively, $A_L^q$ and $A_{1,L}^d$, $A_{2,L}^d$ via relations
analogous to~(\ref{FTqconn}) and~(\ref{FTdconn}) (see
e.~g.,~\cite{MF}).

A component of the truncated scattering matrix $S_{(0,1)}$
in the state with a fixed angular momentum $L$ is diagonal. Its
nontrivial quartet, $s_L^q(z)$, and doublet, $s_L^d(z)$, elements
are given by
$$
   s_L^q(z)=1+2ia_L^q(z),
$$
$$
   s_L^d(z)=1+2ia_L^d(z).
$$
Therefore, the sheet $\Pi_{(01)}$ resonances in the state $L$ of
the $nnp$-system are in fact, zeros of the {\em scalar}
functions $s_L^q(z)$ and $s_L^d(z)$.

When solving the boundary-value problems~(\ref{Fquartet}),
(\ref{BCQ}), (\ref{AsQ}) and~(\ref{Fdoublet}), (\ref{BCD}),
(\ref{AsD}) numerically at complex energies $z$, we use the same
algorithm~\cite{MF}, \cite{MGL}, \cite{EChAYa} as at real $z$.
First, we make a finite-difference approximation of the problems
above in polar coordinates $\rho$, $\theta$.  As the grid points we
take the intersection points of the arcs $\rho=\rho_i$,
$i=1,2,...,N_\rho$, and rays $\theta=\theta_j$,
$j=1,2,...,N_\theta$, $\theta_j < \theta_{j+1}$. At $\rho_0=0$
and given $\rho_1$ (see Table~\ref{tableI}), the grid points in
$\rho$ for $i\geq 2$ are chosen in such a way that
$\rho_{i+1}=\rho_i +{\sf a}(\rho_i-\rho_{i-1})$ with a parameter
${\sf a}$ (acceleration) not depending on the number of a grid
point.  All the results exposed below, are related to ${\sf
a}=1,01.$ The interval $[0,\pi/2]$ where the variable $\theta$
changes, is divided by special points with numbers
$N_\theta^{(k)}$, $k=1,2,3,$ into three subintervals, inside 
which the grid points in $\theta$ are distributed uniformly.
When going from one interval to another (in direction of the
parameter $\theta$ rise), the grid step is divided by two.  The
choice of the grid described, is explained by a necessity to
take into account essentially more quick change of the Faddeev
component values in the domain where $\rho$ and/or $x$~\cite{MF}
are small.  Usually, we chose the numbers of points in $\theta$
and $\rho$ the same, $N_\theta=N_\rho$.  A maximal value of the
parameters $N_\theta$, $N_\rho$ has equaled to 180.  With these
values, the cut-off radius $\rho_{max}=\rho_{N_\rho}$ has
reached 39~fm.  Typical values of the grid parameters are given
in Table~\ref{tableI} where $N=N_\theta=N_\rho$.

\begin{table}
\caption{ Convergence of the $nnp$-system virtual level
$z_{\rm res}({}^3\!{H})$ }
\begin{center}
\begin{tabular}{||c|c|c|c|c|c|c||}
N & $ N_{\theta}^{(1)} $ & $ N_{\theta}^{(2)} $ & $ N_{\theta}^{(3)} $ &
$ \rho_1$, fm &
$ \rho_{max}$, fm & $z_{\rm res}({}^3\!{H}),$ MeV) \\ \hline
 40 & 10 & 20 & 30  & 0.40  & 19.0 & 2.9296 \\
 60 & 15 & 30 & 45  & 0.30  & 23.9 & 2.8229 \\
 80 & 20 & 40 & 60  & 0.25  & 29.9 & 2.7565 \\
120 & 30 & 60 & 90  & 0.15  & 34.0 & 2.7434 \\
160 & 40 & 80 & 120 & 0.10  & 39.1 & 2.7282 \\
180 & 45 & 90 & 135 & 0.08  & 39.5 & 2.7275 \\
\end{tabular}
\end{center}
\label{tableI}
\end{table}

As a nucleon-nucleon interaction, the Malfliet-Tjon potential
MT~I--III is used in its initial version~\cite{MT}.

Having solved the problems~(\ref{Fquartet}), (\ref{BCQ}),
(\ref{AsQ}) and~(\ref{Fdoublet}), (\ref{BCD}), (\ref{AsD})  we
calculate the functions $s_L^q(z)$ and $s_L^d(z)$. Resonances,
considered as roots of these functions in the complex plane, are
found using the Newton method with a three-point approximation
of derivative.

As a test of the computer code we have calculated  the
bound-state energy $E_t$ of the $^3\!{H}$ nucleus as a pole of
the function $s_L^d(z)$. More precisely, this pole was found as
a root of the inverse amplitude $1/a_L^d(z)$.  Beginning from
the grid dimension $80\!\times\! 80$, we obtained for  $E_t$ the
value $-8.55$~MeV.  Hereafter all the energies are given with
respect to the breakup threshold.  Note that the value stated
for $E_t$ is in a good agreement with known results on $E_t$ in
the MT~I--III model (see~\cite{CGM}).

When searching for the $nnp$ resonances on the sheet
$\Pi_{(0,1)}$ at $L=0,1$, we have inspected a domain of a range
about $10$~MeV in vicinity of the segment $[E_d,0]$ in the complex
$z$ plane. Especially carefully we studied a vicinity of the
points $z\! =\! -1.5\pm 0.3+i(0.6\pm 0.3)$~MeV interpreted in
the recent works~\cite{RRCexp}, \cite{RRCBarabanov} as a
location of an exited state energy of the $^3\! H$ nucleus.
Unfortunately we have succeeded to find only one root $z_{\rm
res}({}^3\!{H})$  of the function $s_0^d(z)$, corresponding to
the known virtual state of the $nnp$ system at the total spin {\bf
S}=1/2.  Position of this root for different grids is shown in
Table~\ref{tableI} illustrating a degree of convergence in the
method used.  As one can see from the table, for the maximal of
the grids examined, the grid $180\!\times\! 180$, we have found
{$z_{\rm res}({}^3\!{H}) \! =\! -2.728$~MeV}.  This means that
the calculated virtual level is situated $0.504$~MeV to the left
from the $nd$ threshold $E_d\! =\!  -2.224$~MeV (in the
MT~I--III model~\cite{MT}).  Note that the shift $E_d\! -\!
z_{\rm res}({}^3\!{H})$ found from experimental data on $nd$
scattering, is $0.515$~MeV (see ~\cite{Orlov}).  Its value
computed in a separabilized MT~I--III model on the base of the
momentum space Faddeev equations, equals to
$0.502$~MeV~\cite{Orlov}. As to the resonance $z\! =\! -1.5\pm
0.3+i(0.6\pm 0.3)$~MeV at $L=0$ discussed
in~\cite{RRCexp},~\cite{RRCBarabanov}, it is quite possible to
be situated not on the {\em two-body} sheet $\Pi_{(0,1)}$ but on the
{\em three-body} sheets $\Pi_{(-1,1)}$ or $\Pi_{(1,1)}$
(see~Fig.~\ref{RimSurf-nnp}).  Due to the reasons mentioned at the
beginning of the section, we may unfortunately neither approve
nor disprove this hypothesis. As it should be expected (see the
data on three-nucleon resonances in~\cite{Orlov}), we have
failed to find any resonances in the quartet state at $L=0$ as
well as at $L=1$.

\begin{figure}
\centering
\psfig{file=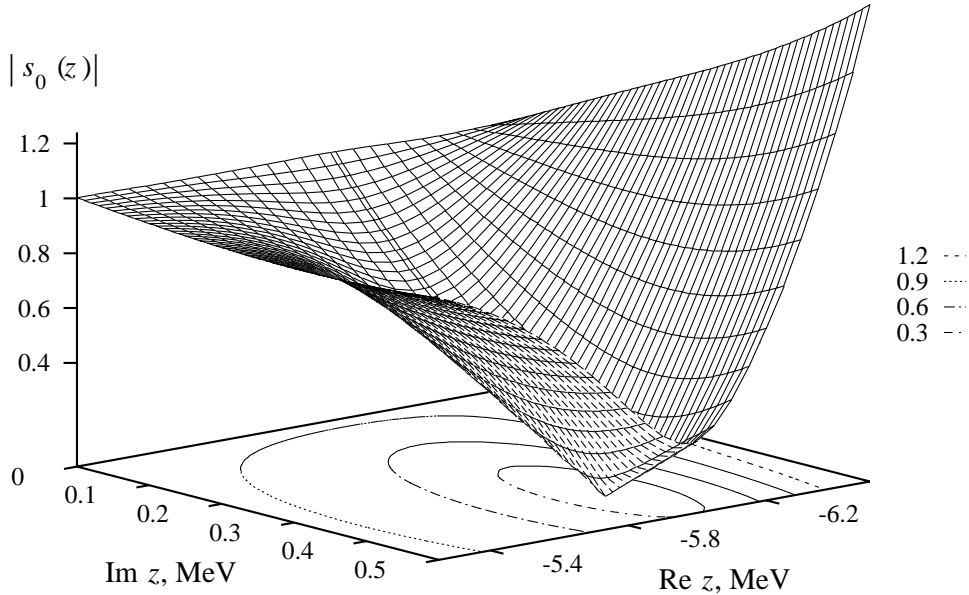,height=10cm}
%
%
%
\caption{ Surface of the function 
$|s_0(z)|$, $s_0(z)\equiv s_0^{\rm 3B}(z)$, in the model
system of three bosons with the nucleon masses.
The potential
$V^G(r)$ is used with the barrier $V_b=1.5$~MeV.
Position of the resonance  $z_{\rm res}(\mbox{3B})$
corresponds to the minimal (zero) value of
$|s_0^{\rm 3B}(z)|$.
}
\label{fig-surface}
\end{figure}
\begin{figure}
\centering
\psfig{file=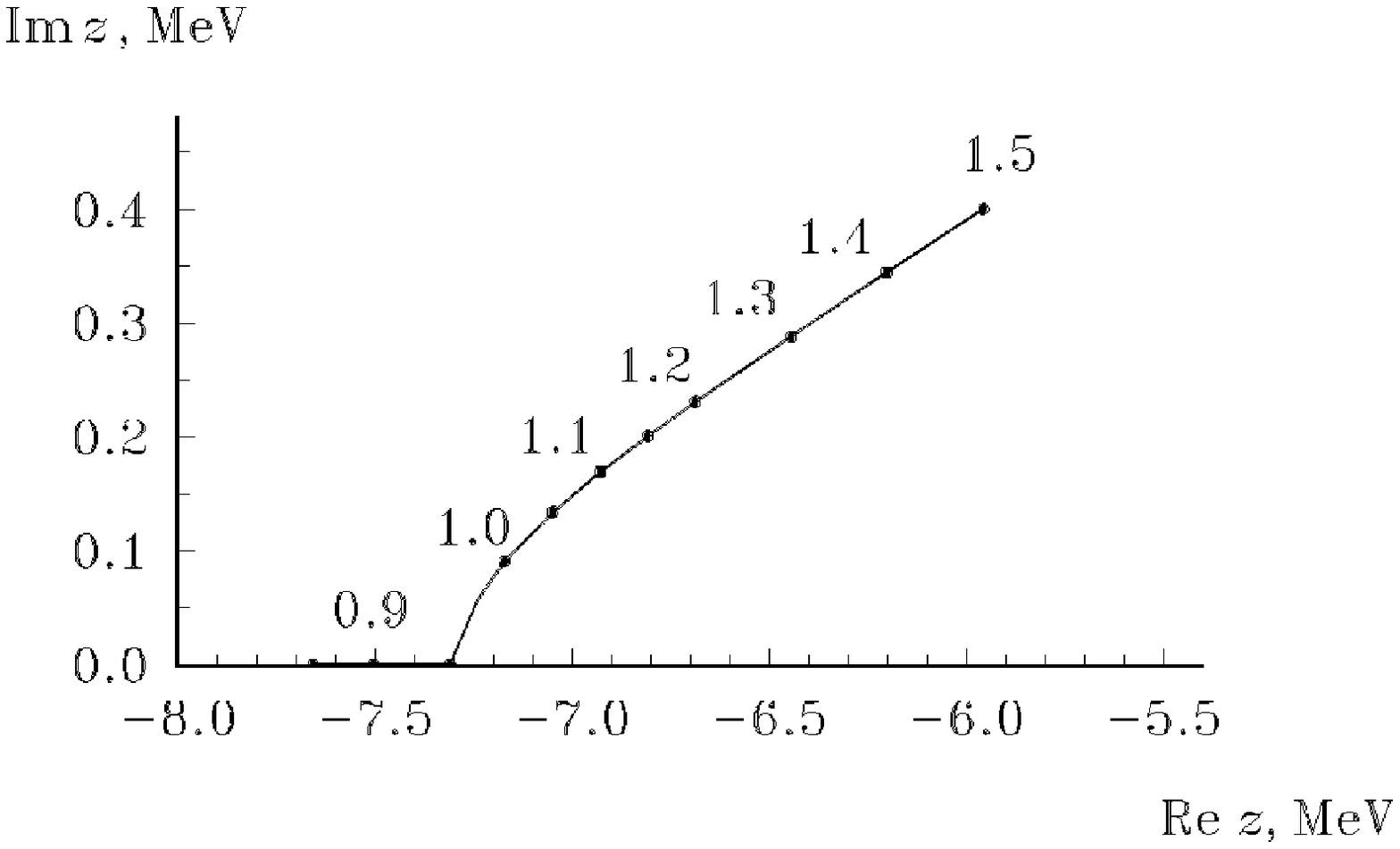,height=10cm}
%
%
%
\caption{
Trajectory of the resonance $z_{\rm res}(\mbox{3B})$ on the 
sheet $\Pi_{(0,1)}$ in the model system of three bosons with the 
nucleon masses. The potential $V^G(r)$ is used. Values of the 
barrier $V_b$ in MeV are given near the points marked on the 
curve.
}
\label{fig-trajectory1}
\end{figure}
\begin{figure}
\centering
\psfig{file=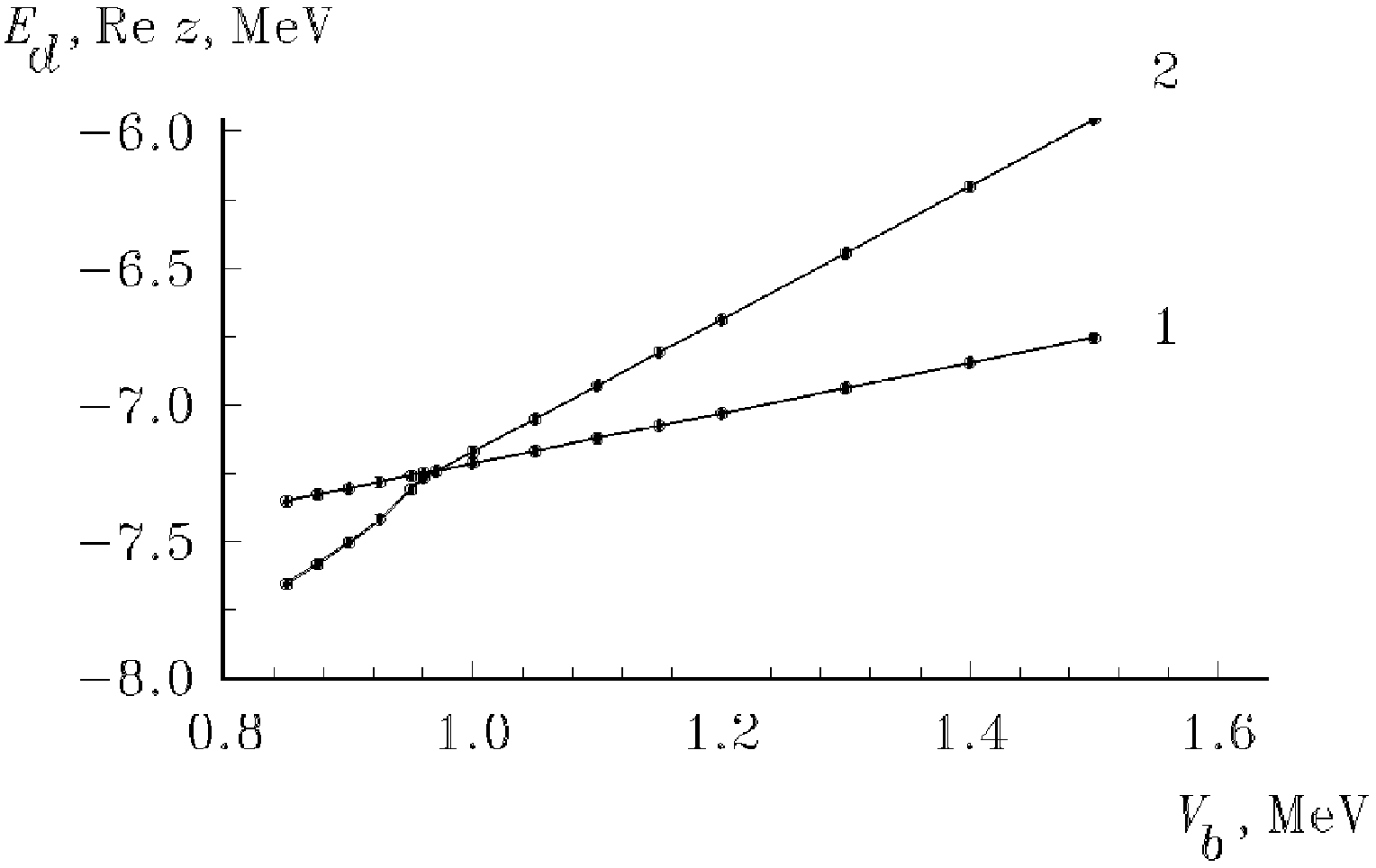,height=10cm}
%
%
%
\caption{
Dependence of the ``deuteron'' energy $E_d$ (curve~1)
and real part of the resonance
$z_{\rm res}(\mbox{3B})$ (curve~2) on the barrier value $V_b$.
}
\label{fig-trajectory2}
\end{figure}

Also, we have studied a model three--body system including
identical spin-zero bosons with masses of the nucleon.  Remind
that the Faddeev integro-differential equations for such
a three-boson system read exactly as the
equations~(\ref{Fquartet}) except a necessity to  replace the
factor $-1/2$ in the expression~(\ref{FTqconn}) with unity.  The
boundary conditions for respective partial Faddeev components
$U^{\rm 3B}_L(x,y)$ have the form~(\ref{BCQ}) and~(\ref{AsQ})
where instead of $a_L^q(z)$ and $A_L^q(\theta,z)$, one has to
substitute $a_L^{\rm 3B}(z)$ and $A_L^{\rm 3B}(\theta,z)$.

Component of the truncated scattering matrix $S_{(0,1)}$ for
the three-boson system is given in the state with the angular
momentum $L$ by
$$
s_L^{\rm 3B}(z)=1+2ia_L^{\rm 3B}(z).
$$
Resonances on the sheet $\Pi_{(0,1)}$ in this state are roots of the
equation $s_L^{\rm 3B}(z)=0$ considered on the physical sheet.

In the three-boson problem we restrict ourselves with a treating
the s-state only and thereby, with a searching for zeros of the
function $s_L^{\rm 3B}(z)$ at $L=0$.  As a pairwise interaction
between the bosons we have used the Gauss-type potential
supplied with an additional Gauss repulsive barrier term,
$$
    V^G(r)=V_0 \exp[-\mu_0 r^2] + V_b \exp[-\mu_b (r-r_b)^2]
$$
where the values $V_0=-55$~MeV, $\mu_0=0.2$~fm$^{-2}$,
$\mu_b=0.01$~fm$^{-2}$,  $r_b=5$~fm have been fixed
and the barrier amplitude
$V_b$ varied.  A resonance (with non-zero imaginary part) on the
sheet $\Pi_{(0,1)}$ arises in the system concerned just due to
the presence of the barrier term.  Example of a surface of the
$s_0^{\rm 3B}(z)$ absolute value for the barrier amplitude
$V_b=1.5$~MeV is shown in Fig.~\ref{fig-surface} (for a
$80\!\times\! 80$~grid).  A trajectory of the resonance $z_{\rm
res}(\mbox{3B})$ (a zero of the function $s_0^{\rm 3B}(z)$) is
shown for the changing barrier $V_b$ in
Fig.~\ref{fig-trajectory1}.  
This trajectory was watched for the
barrier $V_b$ decreasing in the interval between $1.5$~MeV and
$0.85$~MeV.
When drawing the trajectory,  we have used a $160\!\times\!  
160$~grid.  It can be seen from Fig.~\ref{fig-trajectory1} that 
the behavior of the resonance $z_{\rm res}(\mbox{3B})$ is rather 
expected:  with monotonously decreasing real part, the imaginary 
part of the resonance changes also monotonously.In 
Fig.~\ref{fig-trajectory2} we plot both the trajectories of the 
resonance real part $\Real z_{\rm res}(\mbox{3B})$ and two-boson 
binding energy $E_d$.  As one can see from this figure, the 
value of $|\Real z_{\rm res}(\mbox{3B})|$ increases more quickly 
than $|E_d|$, coinciding with $|E_d|$ at $V_b \cong 0.97$~MeV.  

Trajectory of the resonance concerned in the lower complex 
half-plane is symmetric to the curve shown in 
Fig.~\ref{fig-trajectory1} with respect to the real axis.  
Respective points, symmetric to those marked in 
Fig.~\ref{fig-trajectory1}, correspond to the same values of 
$V_b$. For $V_b \approx 0.95$~MeV the energy $z_{\rm 
res}(\mbox{3B})$ becomes real.  In this point the resonances 
$z_{\rm res}(\mbox{3B})$ and $\overline{z}_{\rm res}(\mbox{3B})$ 
``collide'' turning for $V_b < 0.95$~MeV into a couple of 
virtual levels.

\begin{acknowledgements}
The authors are grateful to Professor~V.B.~Belyaev for valuable
discussions and to Dr.~F.M.~Penkov for useful remarks.  
Financial support of the work form the International Science
Foundation and Russian government (grant~RFB300) is kindly 
acknowledged.

\end{acknowledgements}


\newpage
\begin{thebibliography}{99}
\bibitem{Baz}
               {\it Baz~A., Zeldovich~Ya., Perelomov~A.}
               { Scattering, reactions and decays in nonrelativistic
                quantum mechanics.}
               Jerusalem: Israel Program for Scientific Translations,
               1969.%
%
\bibitem{Newton}
         {\it Newton~R.G.}, { Scattering theory of waves and particles.}
        McGraw Hill, New York, 2nd ed., 1982.%
%
\bibitem{AlfaroRegge}
     {\it Alfaro~V.~de, Regge~T.} { Potential scattering}
      [Russian translation].  Mir, Moscow, 1986.%
%
\bibitem{BohmQM}
               {\it B\"{o}hm~A.}
               { Quantum Mechanics: Foundations and Applications.}
                 Springer-Verlag, 1986.%
%
\bibitem{Kukulin}
  {\it Kukulin~V.I., Krasnopol'sky~V.M., Hor\'a\v{c}ek~J.}
  { Theory of resonances: Principles and applications}.
   --- Praha: Academia, 1989.%
%
\bibitem{ReedSimonIII}
     {\it Reed~M., Simon~B.} { Methods of modern mathematical physics.
      III: Scattering theory. }---  N.Y.: Academic Press,  1979.%
%
\bibitem{ReedSimonIV}
     {\it Reed~M., Simon~B.} { Methods of modern mathematical physics.
      IV: Analysis of operators. }---  N.Y.: Academic Press,  1978.%
%
\bibitem{AlbeverioBook}
     {\it Albeverio S., Gesztezy~F., H{\o}egh-Krohn~R., Holden~H.}
     { Solvable models in quantum mechanics}. Springer--Verlag, 1988.%
%
\bibitem{SimonChem}
                    {\it Simon~B.}//
       Intern. J. Quant. Chem. 1978. V.~14. P.~529--542.%
%
\bibitem{GGamow} {\it  Gamow~G.}//
           Z.Phys. 1928. V.~51. P.~204--212.%
%
\bibitem{Jost} {\it  Jost~R.}//
          Helv. Phys. Acta. 1947. V.~20. P.~250--266.%
%
\bibitem{Titchmarsh} {\it Titchmarsh~E.C.}
        { Eigenfunction expansions
      associated with second order differential equations, Vol.~II}.
        --- London: Oxford U.P., 1946.%
%
\bibitem{Orlov}
       {\it M\"{o}ller~K., Orlov~Yu.V.}
       //Fiz. Elem. Chast. At. Yadra.\,
                 1989. V.~20. P.~1341--1395.%
%
\bibitem{BalslevCombes}
         {\it Balslev~E., Combes~J.M.}//
          Commun.Math.Phys. 1971. V.~22. P.~280--294.%
%
\bibitem{Hu} {\it Hu~C.-Y., Bhatia~A.K.}//
             Muon Catalyzed Fusion.
             1990/91. V.~5/6. P.~439--444.%
%
\bibitem{Korobov} {\it Korobov~V.I.}//
      Intern. Sympos. on Muon Catalyzed Fusion, Dubna June~19--24, 1995.
      Book of Abstracts. Dubna: JINR, 1995. P.~77.%
%
\bibitem{Szoto} {\it Cs\'ot\'o~A.}//
         Contributed papers to 14th Intern. Conf. on Few Body
        Problems in Physics. --- Williamsburg: CEBAF,
        1994. P.~777--780.\\
        {\it Cs\'ot\'o~A., Oberhummer~H., Pichler~R.} //
     LANL E-print {\tt nucl-th/9510017}.%
%
\bibitem{Faddeev63}
              {\it Faddeev~L.D.}//  { Trudy Mat. In--ta AN SSSR.}
            1963. V.~69. P.~1--125 [English translation: { Mathematical
               aspects of the three--body problem in quantum mechanics.}
               Israel Program for Scientific Translations,
               Jerusalem, 1965].%
%
\bibitem{MF}
          {\it Faddeev~L.D., Merkuriev~S.P.} { Quantum scattering theory
            for several particle systems.} ---
           Doderecht: Kluwer Academic Publishers, 1993.%
%
\bibitem{MotFewBodyCEBAF} {\it Motovilov~A.K.}//
         Contributed papers to 14th Intern. Conf. on Few Body
        Problems in Physics. --- Williamsburg: CEBAF,
        1994. P.~816--819.%
%
\bibitem{MotRemJINR}
     {\it Motovilov~A.K.} 
//     Theor. Math. Phys. 1996. V.~107. No.~3. P.~784--824.
       (LANL E-prints {\tt nucl-th/9505028} and {\tt nucl-th/9505029}). 
     Math. Nachr.  1997. V.~187. P.~147--210
    (LANL E-print {\tt funct-an/9509003}).
%
\bibitem{MGL}
     {\it Merkuriev~S.P., Gignoux~C., Laverne~A.}//
     Ann. Phys. (N.Y.). 1976. V.~99. P.~30--71.%
%
\bibitem{EChAYa}
      {\it Kvitsinsky~A.A., Kuperin~Yu.A., S.P.Merkuriev~S.P.,
                 Motovilov~A.K., Yakovlev~S.L.}//
                 Fiz. Elem. Chast. At. Yadra.\,
                 1986. V.~17. P.~267--317.%
%
\bibitem{BGC} {\it Benayoun~J.J., Gignoux~C., Chauvin~J.}//
      Phys. Rev.~C. 1981. V.~23. P.~1854--1857.%
%
\bibitem{CPFG}
        {\it  Chen~C.R., Payne~G.L., Friar~J.L., Gibson~B.F.}//
          Phys. Rev.~C. 1989. V.~39. P.~1264--1268;
             Phys. Rev.~C. 1991. V.~44. P.~50-59.%
%
\bibitem{CGM} {\it Carbonell~J., Gignoux~C., Merkuriev~S.P.}//
    Few--Body Systems. 1993. V.~15. P.~15--23.%
%
\bibitem{YakovlevFilikhin}
    {\it Yakovlev~S.L., Filikhin~I.N.}//
     Yad. Fiz. 1993. V.~56. P.~98--106.%
%
\bibitem{TMF93}
   {\it Motovilov~A.K.}//
    Theor. Math. Phys. 1993. V.~95. No.~3. P.~692--699.%
%
\bibitem{MT}
      {\it Malfliet~R.A., Tjon~J.A.}//
       Nucl. Phys.~A. 1969. V.~127. P.~161--168.%
%
\bibitem{RRCexp}
   {\it Alexandrov~D.V., Nikolsky~E.Yu., Novatsky~B.G.,
        Stepanov~D.N.}//
        Pis'ma ZhETF. 1994. V.~59. P.~301--304.%
%
\bibitem{RRCBarabanov}
     {\it Barabanov~A.L.}
     // Pis'ma v ZhETF. 1995. V.~61. P.~9--14 
       (LANL E-print {\tt nucl-th/9412036}).%
\end{thebibliography}
\end{document}